\documentclass[preprint,superscriptaddress,frontmatterverbose,aps,prd,amsmath,amssymb,floatfix,notitlepage,nofootinbib]{revtex4-2}

\usepackage{graphicx}
\usepackage{hyperref}
\usepackage{orcidlink}
\usepackage{hyperref}

\usepackage{enumitem}
\setlist{nosep, topsep=\parskip}

\def \beq{\begin{equation}}
\def \eeq{\end{equation}}
\def \bea{\begin{eqnarray}}
\def \eea{\end{eqnarray}}

\def \({\left(}
\def \){\right)}
\def \[{\left[}
\def \]{\right]}

\def\btopik{B\to\pi K}

\allowdisplaybreaks

\begin{document}

\title{\boldmath Comment on ``QCD-factorization amplitudes from flavour symmetries: beyond the $SU(3)$ symmetric case''}

\author{Bhubanjyoti Bhattacharya\,\orcidlink{0000-0003-2238-321X},}
\email{bbhattach@ltu.edu}
\affiliation{Department of Natural Sciences, Lawrence Technological University, Southfield, MI 48075, USA}

\author{and David London\,\orcidlink{0000-0002-4407-5624}}
\email{london@lps.umontreal.ca}
\affiliation{Physique des Particules, Universit\'e de Montr\'eal, 1375 Avenue Th\'er\`ese-Lavoie-Roux, Montr\'eal, QC, Canada  H2V 0B3}

\date{\today}

\preprint{UdeM-GPP-TH-26-309}

\begin{abstract}
Recently, a fit to $B \to PP$ decays ($P \in \{\pi, K, \eta, \eta'\}$) was performed (\href{https://arxiv.org/abs/2604.19612}{``QCD-factorization amplitudes from flavour symmetries: beyond the $SU(3)$ symmetric case''}
) using a formalism that combines topological diagrams with QCD factorization, and a good fit was found. We also recently performed such a fit, under the assumption that the $B \to PP$ amplitudes are related by flavour SU(3) symmetry, but we found a very poor fit. The two results therefore disagree with one another. The source of this disagreement is that we applied EWP-tree relations (ETRs). These were derived $\sim 30$ years ago, and relate different topological diagrams or reduced matrix elements, thus reducing the number of unknown parameters in the fit. In their paper, it is asserted that ETRs are invalid, so that analyses that use them are unreliable. We are writing this Comment to explain why this assertion is incorrect. The key point is that ETRs are mathematically rigorous, group theoretically. If SU(3) is unbroken, and the small Wilson coefficients $c_{7,8}$ in the weak effective Hamiltonian are neglected, ETRs follow automatically and are exact. That is, this is a group theory result -- no hadronic calculations are involved. In this Comment, we also point out several weaknesses of their formalism.
\end{abstract}

\maketitle
\newpage

In Ref.~\cite{Neubert:1998pt}, Neubert and Rosner established a relation between the electroweak penguin (EWP) and tree topological diagrams of Ref.~\cite{Gronau:1994rj}. Following this, in Ref.~\cite{Gronau:1998fn}, Gronau, Pirjol, and Yan rederived the Neubert-Rosner EWP-tree relation (ETR), and derived another, previously unknown ETR. They used a rigorous group-theoretical method relating the reduced matrix elements (RMEs) of flavour SU(3) symmetry to topological diagrams. Both the Neubert-Rosner and Gronau-Pirjol-Yan ETRs apply to charmless $B \to PP$ decays. Two assumptions were made in deriving these relations: (i) the decays are related by unbroken SU(3), and (ii) the Wilson coefficients $c_{7,8}$ are neglected. 

In Refs.~\cite{Berthiaume:2023kmp, Bhattacharya:2025wcq}, with collaborators, we performed a fit to charmless $B\to PP$ decays under the same assumptions. A very poor fit was found. Later, in Ref.~\cite{Bhattacharya:2025rrv}, we rederived the SU(3) ETRs, this time purely at the level of group theory using SU(3) RMEs. At the same time, we derived ETRs with isospin as the underlying symmetry, and demonstrated that these \emph{new} isospin-based ETRs are different from the previously-known SU(3)-based ETRs. 

The key point here is that these results are completely rigorous. As long as the underlying symmetry is unbroken, and $c_{7,8}$ are neglected, the ETRs are exact. And if $c_{7,8}$ are not neglected, we showed in Ref.~\cite{Bhattacharya:2025rrv} that the ETRs are modified by only $\mathcal{O}({10\%})$. Note that, since isospin is an almost exact symmetry of QCD, large isospin-breaking effects are not expected in the SM. This implies that large violations of the isospin-based ETRs are also unexpected. Recently, following our analysis, Ref.~\cite{Grossman:2026qew} quantified the effect of isospin breaking in the direct CP asymmetry of $B^+\to\pi^+\pi^0$.

In ``QCD-factorization amplitudes from flavour symmetries: beyond the $SU(3)$ symmetric case,'' by W.-S. Fang, T. Huber, X.-Q. Li, E. Malami and G. Tetlalmatzi-Xolocotzi \cite{Fang:2026fhl}, a fit to charmless $B \to PP$ decays was performed using a different formalism than ours, and an acceptable fit is found. The problem is that this paper contains numerous incorrect and misleading statements about ETRs and our analysis. We are writing this Comment to set the record straight.

As discussed above, ETRs have been around for almost 30 years, and they are rigorous mathematical results, using only group theory. Despite this, the authors of Ref.~\cite{Fang:2026fhl} take great pains to argue that there are problems with ETRs. The main reason is that, when ETRs are taken into account, the number of independent SU(3) RMEs in $B \to PP$ decays is reduced from 10 to 7 in the SU(3) limit. This is what was used in our fits of Refs.~\cite{Berthiaume:2023kmp, Bhattacharya:2025wcq}. On the other hand, in Ref.~\cite{Fang:2026fhl}, they take all 10 RMEs to be independent. With more unknown parameters, it is not surprising that they find a good fit. To justify the use of 10 independent parameters in their fit, they claim that ETRs are unreliable. 

For example, in the introduction, the authors of Ref.~\cite{Fang:2026fhl} write\footnote{Note that, when a reference appears within a quotation in this Comment, we use the reference number as it appears in our list of references, rather than that in the source. Thus, Ref.~[99] of Ref.~\cite{Fang:2026fhl} is cited here as Ref.~\cite{Shi:2025eyp} even when it appears within a quotation.} ``We argue that these relations do not withstand quantum or power corrections (see the derivation in \cite{Shi:2025eyp}), and therefore advocate to use the complete set of independent amplitudes for phenomenology.'' The problem here is that Ref.~\cite{Shi:2025eyp} says nothing of the sort -- there is no derivation that would signify any quantum or power corrections. On the contrary, the authors of Ref.~\cite{Shi:2025eyp} agree with us: in their Sec.~IV, they derive the ETRs in terms of their topological diagrams. They conclude with ``The three RMEs that are neglected in Ref.~\cite{Berthiaume:2023kmp} arise from the contributions from the electroweak penguin operators $O_{7,8}$. Since the Wilson coefficients for $O_{7,8}$ are small in the standard model, neglecting these contribution may not have sizable impact in phenomenological analysis.'' Since the topological diagrams of Ref.~\cite{Shi:2025eyp} are the same as those of Ref.~\cite{Fang:2026fhl}, there is no justification for asserting that ETRs are unreliable. In fact, following what was actually said in Ref.~\cite{Shi:2025eyp}, these ETRs should have been used in the analysis of Ref.~\cite{Fang:2026fhl}.

There is more. In Sec.~6.4 of Ref.~\cite{Fang:2026fhl}, ``Scrutinizing the EWP–tree relations,'' it is stated that ``the EWP–tree relations can be derived under the strict leading-order (LO) QCDF limit, with two non-trivial implicit assumptions: (i) all non-factorizable QCD corrections, including vertex, penguin, hard-spectator scattering, and weak annihilation contributions, are neglected; (ii) the strong-interaction dynamics of EWP amplitudes are identical to those of tree amplitudes, such that their hadronic matrix elements share a common factorizable structure.'' But look at the ETRs derived in this way, their Eq.~(6.10). These are the isospin-based ETRs! This is very strange. Did they consider all $B \to PP$ decays under the assumption of SU(3) and somehow derive the isospin-based ETRs? Or did they consider only $\btopik$ decays with isospin symmetry, and forget to tell us? Whatever the answer, this puts into doubt this derivation, and raises serious questions about the research in this paper in general.

Let us take this result at face value, i.e., they are somehow able to derive the isospin-based ETRs within QCDF. They then go on to say that, if the ratios of diagrams are computed using the favoured values of the diagrams found in their fit, these disagree by a factor of 20-30 with the ratios as computed using the ETRs. They suggest that this implies that there is something wrong with the ETRs. There are two obvious problems with this assertion. First, their fit was done assuming SU(3), not isospin, so they should be comparing their results with the SU(3)-based ETRs. Second, why does this imply that there is a problem with the ETRs? This really appears to be a problem with the QCDF framework used in Ref.~\cite{Fang:2026fhl}.

But they go further: they say that, when nonfactorizable corrections are included in the QCDF calculation, the ETRs are badly broken. But these are the isospin-based ETRs. As explained previously, since isospin is very close to an exact symmetry of QCD, we expect the isospin-based ETRs to be almost exact. If QCDF is unable to reproduce these relations, this is a problem of QCDF. To be fair, it is probably a problem of the way that the authors of Ref.~\cite{Fang:2026fhl} are doing their QCDF calculation. But once again, this puts into doubt their entire approach.

They conclude this section with, ``Our findings clearly demonstrate that the naive EWP–tree relations are badly broken, rendering the simple proportionality to tree amplitudes invalid, and reinforce the necessity of treating the EWP amplitudes as fully independent parameters in the TDA analyses \cite{He:2018php, Shi:2025eyp, Beneketalk}. Consequently, any analysis that imposes the EWP–tree relations a priori runs the risk of omitting the dominant dynamical component of these amplitudes.'' This is demonstrably false. First, the ``naive EWP–tree relations'' that they refer to are derived using only symmetry considerations (i.e., group theory), and are mathematically rigorous. They are not the result of a hadronic calculation. Second, these same ETRs were derived in the TDA formalism in Ref.~\cite{Shi:2025eyp}. Finally, it appears that the authors of Ref.~\cite{Fang:2026fhl} cannot reproduce the (almost exact) isospin-based ETRs. All of this shows quite clearly that ETRs are not the problem. Rather, there is something wrong with the methodology of Ref.~\cite{Fang:2026fhl}.

The point of the above discussion is that, contrary to what is stated repeatedly in Ref.~\cite{Fang:2026fhl}, there is no problem with ETRs. As long as SU(3) is unbroken, the ETRs relate topological diagrams or RMEs that are a-priori independent. This said, Ref.~\cite{Fang:2026fhl} claims to go ``beyond the $SU(3)$ symmetric case.'' Although they work with the standard weak effective Hamiltonian [their Eq.~(2.1)], the number of free parameters they use (18 RMEs when $\eta$ and $\eta'$ are included in the analysis) is greater than 10, the number of independent RMEs that one would find when ETRs are ignored. They give the following ``justification:'' ``Note that in the exact $SU(3)$ flavour symmetry limit, these topological amplitudes are symmetric under the permutation of the final-state mesons. As the symmetry is already broken by the splitting between the up/down and strange quark masses, here we do not treat them as symmetric but rather consider them to be dependent on the ordering of the final-state mesons $M_1$ and $M_2$, which provides us an easy way to quantify the amount of flavour-$SU(3)$ breaking effects in terms of transition form factors and decay constants.'' In other words, their entire analysis is done assuming that SU(3) is broken. It is therefore meaningless for them to compare their results with ours, which are model-independent fits based on an SU(3)-symmetric parametrization.

Let us look more closely at the above ``justification.'' The key question is how to implement SU(3) symmetry and SU(3) breaking in the analysis of $B\to PP$ decays. The correct way of doing this is to begin with SU(3)-symmetric amplitudes and then to add a symmetry-breaking spurion in the form of a strange-quark mass term. In this way, one has the ability to treat SU(3)-breaking contributions perturbatively, as one should when the SU(3)-breaking effects are small. (As an example of this, Ref.~\cite{Gronau:2013mda} examines two-body $B$ decays in the context of U-spin symmetry and U-spin breaking.) In particular, one should not just assume that the final state depends on the ordering of mesons. Rather, one should consider the effect of applying the spurion to the effective Hamiltonian to create a breaking Hamiltonian. 

The point is that what is done in Ref.~\cite{Fang:2026fhl} cannot be characterized as going ``beyond the $SU(3)$ symmetric case'' since the final states they begin with are not SU(3) symmetric. A true analysis of SU(3) breaking would begin with an SU(3)-symmetric theory, including the ETR relations, and then systematically add SU(3)-breaking effects. What they have done is to add some model-dependent QCDF corrections, but this is in no way an SU(3) analysis. In fact, since their approach does not have a mechanism to adequately restore the full SU(3) symmetry, it is not surprising that they are unable to reproduce the ETRs, which are derived under the assumption of unbroken SU(3) symmetry.

Finally, we must mention the $B \to K \pi$ sum rule, discussed in Sec.~6.3 of Ref.~\cite{Fang:2026fhl}. They state that, in the SM, it is expected that $\Delta_{\rm SR} = 0$. The irony here is amusing. This sum rule was first derived in Ref.~\cite{Gronau:2005kz}. There it was argued that, if one uses the SU(3) ETRs and adds some theory input, one finds $\Delta_{\rm SR} \approx 0$. However, it was shown in Ref.~\cite{Bhattacharya:2025rrv} that, if one uses instead the isospin-based ETRs, which are appropriate when considering $\btopik$ decays, one finds that the sum rule is no longer approximate: $\Delta_{\rm SR} = 0$ is predicted. In other words, while they stress the importance of this sum rule, the fact is that it is a {\it consequence} of the ETRs! Note also that they find $\Delta_{\rm SR} = -0.09 \pm 0.03$ in their fit. Despite differing from 0 by $3\sigma$, they say this is compatible with the SM expectation.

To summarize,
\begin{enumerate}

\item The claim that ``the naive EWP–tree relations are badly broken'' is incorrect. If SU(3) is unbroken and the Wilson coefficients $c_{7,8}$ are neglected, the ETRs can be derived straightforwardly and are exact. That is, they are rigorous mathematical results, based only on group theory. 

\item When one uses the ETRs, the number of independent SU(3) RMEs is reduced from 10 to 7. The authors argue that, because the ETRs are badly broken, it is necessary to keep all 10 RMEs as independent in the TDA analysis. The problem is that Ref.~\cite{Shi:2025eyp}, cited in their paper, does not agree with this. Ref.~\cite{Shi:2025eyp} works in the TDA formalism, derives ETRs in terms of their topological diagrams, and says this leads to 7 RMEs. In other words, if one does a TDA analysis, the ETRs should be imposed and one should use 7 independent RMEs.

\item The authors state that ETRs can be derived within QCDF, but this requires some very non-trivial assumptions. Furthermore, when nonfactorizable corrections are included in the calculation, the ETRs are badly broken. They conclude that there is a problem with the ETRs. This is incorrect -- the problem is not with the ETRs, but with the QCDF calculation. First, for some reason, the ETRs they derive are not the SU(3)-based ETRs, but the isospin-based ETRs. Second, the isospin-based ETRs are almost exact -- isospin-breaking effects are very small. The fact that QCDF is unable to reproduce these relations shows that there is a problem with the calculation. And this calls into question the entire formalism of the paper.

\item The authors say that their formalism goes “beyond the $SU(3)$-symmetric case.” In our opinion, their formalism cannot be characterized in this way. In order to properly analyze SU(3) breaking, it is necessary to begin with an SU(3)-symmetric theory (including the ETRs) and then systematically add SU(3)-breaking effects. This is not what they have done. At the beginning of their calculation, they added some model-dependent QCDF corrections. However, this is not an SU(3) analysis since one cannot restore the full SU(3) symmetry in their formalism.

\item The authors stress the importance of the $B \to K \pi$ sum rule, which says that $\Delta_{\rm SR} = 0$ is predicted in the SM. But the fact is that this prediction is a {\it consequence} of the ETRs, those same ETRs that they claim are unreliable.

\end{enumerate}
    
\acknowledgments{We thank Marianne Bouchard, Alexandre Jean, and Ipsita Ray for discussions. This work was financially supported by the National Science Foundation, Grant No.\ PHY-2310627 (B.B.) and by NSERC of Canada (D.L.)}

\bibliography{comment}
\bibliographystyle{apsrev4-2}

\end{document}